\documentclass[twocolumn,pra,showpacs,superscriptaddress]{revtex4}
\usepackage{amssymb}
\usepackage{amsmath}
\usepackage{graphicx}
\usepackage{subfigure}
\usepackage{natbib}
\usepackage{epsfig}
\usepackage{amsfonts}
\usepackage{mathrsfs}
\usepackage{ulem}
\usepackage{color}
\usepackage[toc,page,title,titletoc,header]{appendix}

\begin{document}

\title{Modified Bethe-Peierls boundary condition for ultracold atoms with
Spin-Orbit coupling}
\author{Peng Zhang}
\email{pengzhang@ruc.edu.cn}
\affiliation{Department of Physics, Renmin University of China, Beijing, 100190, China}
\author{Long Zhang}
\affiliation{Department of Physics, Renmin University of China, Beijing, 100190, China}
\affiliation{Hefei National Laboratory for Physical Sciences at Microscale and Department
of Modern Physics, University of Science and Technology of China, Hefei,
Anhui 230026, China}
\author{Youjin Deng}
\affiliation{Hefei National Laboratory for Physical Sciences at Microscale and Department
of Modern Physics, University of Science and Technology of China, Hefei,
Anhui 230026, China}

\begin{abstract}
We show that the Bethe-Peierls (BP) boundary condition should be modified
for ultracold atoms with spin-orbit (SO) coupling. Moreover, we derive a
general form of the modified BP boundary condition, which is applicable to a
system with arbitrary kind of SO coupling. In the modified BP condition, an
anisotropic term appears and the inter-atomic scattering length is normally
SO-coupling dependent. For the special system in the current experiments,
however, it can be proved that the scattering length is SO-coupling
independent, and takes the same value as in the case without SO coupling.
Our result is helpful for the study of both few-body and many-body physics
in SO-coupled ultracold gases.
\end{abstract}


\maketitle

\section{Introduction}

In the study of ultracold atomic gases~\cite%
{fermireview,fewbodypetrov,castin2,aad,add,castin1,hu1,hu2,hu3}, the
Bethe-Peierls (BP) boundary condition~\cite{bpc} is widely used as a
replacement of realistic interaction between two atoms.
%of ultracold atomic gases~\cite%
%{fermireview,fewbodypetrov,castin2,aad,add,castin1,hu1,hu2,hu3}. In
%theoretical calculations, the BP boundary condition can be used as a
%replacement of the realistic interaction potential between two atoms.
With this approach, one only needs to solve the Schr\"{o}dinger equation
with the Hamiltonian for atomic free motion, and thus the calculation is
significantly simplified. As a result, the BP boundary condition is very
useful in the research of few-body and many-body physics in ultracold gases,
especially those with large interatomic scattering lengths. Many
achievements have been obtained. For instance, for two-component Fermi
gases,
%especially when interatomic scattering lengths are large, e.g., the two-component
%Fermi gases in the unitary region.
%using the BP boundary condition.  For instance,
Petrov \emph{et. al.} obtained the atom-dimer~\cite{aad} and the dimer-dimer~%
\cite{add} scattering lengths, and Werner \emph{et.al.}~\cite{castin1}
rederived the well knonwn Tan's relations~\cite{tan1,tan2,tan3} using the BP
boundary condition. %for the two-component fermionic gas. Recently,
%of the same system can also be obtained with the BP
%boundary condition~\cite{castin1}.

In recent years, a class of synthetic gauge fields and spin-orbit (SO)
coupling has been realized in ultracold Bose gases~\cite%
{NIST,NIST_elec,SOC,collective_SOC,NIST_partial,JingPRA,ourdecay} and
degenerate Fermi gases~\cite{SOC_Fermi,SOC_MIT}. A considerable amount of
theoretical interest has been stimulated to understand the SO-coupling
effect in both few-body~\cite{fewbody1,fewbody2,fewbody3,fewbody4,fewbody5}
and many-body physics~\cite%
{review,Stripe,Ho,Wu,Victor,Hu,Santos,vyasanakere-11, gong-11, yu-11, hu-11,
iskin-11, yi-11, han-12, dellanna-11, zhou-11b, chen-12, seo-11, huang-11,
he-11,m-2}, including the gases with large interatomic scattering lengths~%
\cite{vyasanakere-11, gong-11, yu-11, hu-11, iskin-11, yi-11, han-12,
dellanna-11, seo-11,m-2}. It becomes now an urgent task to carefully examine
the BP boundary condition in SO-coupled ultracold gases.

In this paper we show that in SO-coupled systems, the BP boundary condition
should be modified and moreover derive a general form of the modified BP
boundary condition that is applicable to a system with any kind of SO
coupling and arbitrary atomic spin. The relevance to the current experiments
is discussed.

%Our work would help the study of
%few-body and many-body physics of the SO-coupled ultracold gases.

%The remainder of this manuscript is organized as follows.
%In Sec. II, the modified BP boundary condition for a spin-$1/2$ fermonic gas,
%and it is generalized to systems
%with arbitrary spin. A brief discussion is given in Sec. IV.
%In Sec. III the modified BP boundary condition for the gases of atoms

\section{Modified BP boundary condition for spin-$1/2$ fermionic atoms}

In this section we shall consider a system of two spin-$1/2$ fermonic atoms
with a short-ranged and spin-dependent interaction potential $U(\vec{r})$,
where $\vec{r}=(x,y,z)$ is the relative position of the two atoms. The
interaction $U(\vec{r})$ has an effective range $r_{\ast }$ such that $U\left(
\vec{r}\right) \simeq 0$ for $r\equiv |\vec{r}|\gtrsim r_{\ast }$.
Furthermore, we shall focus on the case of low-energy scattering for which
the difference $\varepsilon $ between the energy of atomic relative motion
and the scattering threshold is much smaller than $1/r_{\ast }^{2}$.

Let $|\!\!\uparrow \rangle $ and $|\!\!\downarrow \rangle $ represent the
spin eigen-states of a single atom, the quantum state of the relative atomic
motion can be described by a spinor wave function $|\psi (\vec{r})\rangle $:
\begin{equation}
|\psi (\vec{r})\rangle =\psi _{S}\left( \vec{r}\right) |{\mathrm{S}}\rangle
+\sum_{j=1}^{3}\psi _{T_{j}}\left( \vec{r}\right) |{\mathrm{T}}_{j}\rangle
\,,
\end{equation}%
where $|{\mathrm{S}}\rangle =\left( |\!\!\uparrow \rangle
_{1}|\!\!\downarrow \rangle _{2}-|\!\!\downarrow \rangle _{1}|\!\!\uparrow
\rangle _{2}\right) /\sqrt{2}$ is the singlet spin state and $|{\mathrm{T}}%
_{j}\rangle $ ($j=1,2,3$) are the three triplet states. In dilute ultracold
gases, the interatomic distance is much larger than the effective range $%
r_{\ast }$, and the physical property of the system is determined by the
behavior of the wave function $|\psi (\vec{r})\rangle $ in the region $%
r\gtrsim r_{\ast }$.

Our task is to investigate the behavior of the wave function $|\psi(\vec{r}%
)\rangle$ in the presence of SO coupling, and then establish the correct BP
boundary condition.

\subsection{Without SO coupling}

For completeness, we start with the case without SO coupling, for which the
relative motion of the two atoms is governed by the Hamiltonian
\begin{equation}
H=\vec{p}^{\,2}+B+U(\vec{r})\,,  \label{2}
\end{equation}%
where $\vec{p}=-i\nabla $ is the relative momentum and the natural units $%
\hbar =m=1$ ($m$ is the single-atom mass) are used. Operator $B$ acts in the
spin space and accounts for the possible $\vec{r}$-independent contribution,
e.g., from the Zeeman effect. We assume the difference of the eigen-energies
of $B$ are much smaller than $1/r_{\ast }^{2}$.

We first consider the property of $|\psi (\vec{r})\rangle $ in the
short-range region $r_{\ast }\lesssim r<<1/\sqrt{\varepsilon }$. According
to the low-energy scattering theory (appendix A), when $|\psi (\vec{r})\rangle $ is a
low-energy eigenfunction of $H$, one has
\begin{equation}
|\psi (\vec{r})\rangle \propto \left( \frac{1}{r}-\frac{1}{a}\right) |%
\mathrm{S}\rangle \hspace{5mm}\mbox{for }r_{\ast }\lesssim r<<1/\sqrt{%
\varepsilon }\,,  \label{bps}
\end{equation}%
with the scattering length $a$ being determined by the detail of $U(\vec{r})$%
. Note that Eq.~(\ref{bps}) does not depend on the eigen-value of $H$ for $%
|\psi (\vec{r})\rangle $, and is thus applicable to to all low-energy wave
functions.

With Eq.~(\ref{bps}), one can obtain the behavior of a low-energy wave
function $|\psi (\vec{r})\rangle $ in the whole region $r\gtrsim r_{\ast }$.
Let $|\phi (\vec{r})\rangle $ be the solution of the Schr\"{o}dinger
equation with Hamiltonian $\vec{p}^{\,2}+B$, together with the BP boundary
condition
\begin{equation}
\lim_{r\rightarrow 0}|\phi (\vec{r})\rangle \propto \left( \frac{1}{r}-\frac{%
1}{a}\right) |\mathrm{S}\rangle +\mathcal{O}(r)\,,  \label{bps2}
\end{equation}%
the realistic wave function $|\psi (\vec{r})\rangle $ and the pseudo one $%
|\phi (\vec{r})\rangle $ will have the same behavior for $r\gtrsim r_{\ast }$%
. Therefore, in the study of few-body and many-body physics in systems with
interaction $U(\vec{r})$, one can replace the realistic potential $U(\vec{r}%
) $ by the BP condition (\ref{bps2}). Theoretical calculations can be
greatly simplified.

\subsection{With one-dimensional SO coupling}

We now consider a simple case with one-dimensional SO coupling. Without loss
of generality, the single-atom Hamiltonian of the system can be written as
\begin{equation}
H_{\mathrm{1b}}=\frac{\vec{P}^{2}}{2}+\lambda \hat{\sigma}_{z}P_{x}+Z\,,
\label{h1b2}
\end{equation}%
where $\vec{P}$ is the atomic momentum, $\hat{\sigma}$ is the Pauli
operator, $\lambda $ indicates the intensity of the SO coupling, and $Z$
accounts for the residual spin-dependent part. This paper shall consider the
case of weak SO coupling $\lambda <<1/r_{\ast }$, as in the current
experiments. We assume the difference between the eigenvalues of $Z$ is much
smaller than $1/r_{\ast }^{2}$.

The total Hamiltonian of the two atoms is given by $H_{\mathrm{1b}}^{(1)}+H_{%
\mathrm{1b}}^{(2)}+U(\vec{r})$, where $H_{\mathrm{1b}}^{(i)}$ is for the $i$%
th atom. Because the total momentum of the two atoms is conserved, the
relative motion can be separated from the mass-center motion. The
Hamiltonian for the relative motion is then%
\begin{equation}
H={\vec{p}}^{\,2}+\lambda \left( \hat{\sigma}_{z}^{(1)}-\hat{\sigma}%
_{z}^{(2)}\right) {p}_{x}+B(\vec{K})+U(\vec{r})\equiv H_{0}+U(\vec{r})\,,
\label{hh}
\end{equation}%
where $B(\vec{K})=Z^{(1)}+Z^{(2)}+\lambda (\hat{\sigma}_{z}^{(1)}+\hat{\sigma%
}_{z}^{(2)})K_{x}{/2}$ and the c-number $\vec{K}=(K_{x},K_{y},K_{z})$ is
just the total momentum of the two atoms.

For the aim of establishing a correct BP boundary condition, one should also
examine the behavior of the eigenfunction $|\psi (\vec{r})\rangle $ of $H$
in Eq.~(\ref{hh}) in the short-range region, now defined as $r_{\ast
}\lesssim r<<r_{s}$ with $r_{s}=\min (1/\sqrt{\varepsilon },1/\lambda )$.
Comparing Eq.~(\ref{hh}) to Eq.~(\ref{2}), one finds that due to the SO
coupling, the Hamiltonian $H$ is modified by a term $\lambda (\hat{\sigma}%
_{z}^{(1)}-\hat{\sigma}_{z}^{(2)}){p}_{x}$. This term exists in the whole
range of the interatomic distance $r$, including the short-range region and
the region $r\lesssim r_{\ast }$. Therefore, the short-range behavior of $%
|\psi (\vec{r})\rangle $ can no longer be described by Eq.~(\ref{bps}), and
the BP boundary condition in Eq.~(\ref{bps2}) cannot be directly applied.

To overcome this difficulty, we introduce a unitary transformation
(rotation) $\mathcal{R}(\vec{r})$ as
\begin{equation}
\mathcal{R}(\vec{r})=e^{i\lambda \left( \hat{\sigma}_{z}^{(1)}-\hat{\sigma}%
_{z}^{(2)}\right) x/2},  \label{r1d}
\end{equation}%
with $x$ the relative position in the $x$ direction, and define the rotated
wave function $|\psi (\vec{r})\rangle _{\mathrm{R}}$ as
\begin{equation}
|\psi (\vec{r})\rangle _{\mathrm{R}}=\mathcal{R}(\vec{r})|\psi (\vec{r}%
)\rangle .
\end{equation}%
An immediate observation is that since $|\psi ({\vec{r}})\rangle $ is an
eigenfunction of Hamiltonian $H$, the rotated wave function $|\psi ({\vec{r}}%
)\rangle _{\mathrm{R}}$ is an eigenfunction of the rotated Hamiltonian
\begin{equation}
H_{\mathrm{R}}=\mathcal{R}(\vec{r})H\mathcal{R}^{\dagger }(\vec{r}).
\end{equation}%
Straightforward calculations yield
\begin{equation}
H_{\mathrm{R}}={\vec{p}}^{\,2}+W(\vec{r})+U_{\mathrm{R}}(\vec{r})
\label{hrs}
\end{equation}%
with
\begin{eqnarray}
U_{\mathrm{R}}(\vec{r}) &=&\mathcal{R}(\vec{r})U(\vec{r})\mathcal{R}%
^{\dagger }(\vec{r})\,,  \label{ur} \\
W(\vec{r}) &=&\mathcal{R}(\vec{r})B(\vec{K})\mathcal{R}^{\dagger }(\vec{r}) -%
\frac{\lambda ^{2}}{4}(\hat{\sigma}_{z}^{(1)}-\hat{\sigma}_{z}^{(2)})^{2}\,.
\end{eqnarray}

Equation~(\ref{hrs}) shows that the SO coupling \textit{disappears} in the
rotated Hamiltonian $H_{\mathrm{R}}$. Furthermore, in the region $%
r<<1/\lambda $, we have $W(\vec{r})\simeq W(0)$, and then
\begin{equation}
H_{\mathrm{R}}\approx H_{\mathrm{SR}}\equiv \vec{p}^{\,2}+W(0)+U_{\mathrm{R}%
}(\vec{r}).  \label{hsr1}
\end{equation}%
Thus, the eigenfunctions of $H_{\mathrm{R}}$ and $H_{\mathrm{SR}}$ have the
same behavior in the short-range region $r_{\ast }<r<<r_{s}$. Further, Eq. (%
\ref{hsr1}) has the same as Eq.~(\ref{2}). Therefore, the eigenfunction $%
|\psi (\vec{r})\rangle _{\mathrm{R}}$ behaves as
\begin{equation}
|\psi (\vec{r})\rangle _{\mathrm{R}}\propto \left( \frac{1}{r}-\frac{1}{a_{%
\mathrm{\scriptscriptstyle R}}}\right) |\mathrm{S}\rangle \hspace{3mm}%
\mbox{for }r_{\ast }\lesssim r<<r_{s},  \label{kakab}
\end{equation}%
analogous to Eq.~(\ref{bps}). It should be pointed out that $a_{\mathrm{%
\scriptscriptstyle R}}$ is the scattering length with respect to the rotated
interaction potential $U_{\mathrm{R}}(\vec{r})$. On the basis of Eq.~(\ref%
{kakab}), the behavior of the wave function $|\psi (\vec{r})\rangle $ in the unrotated frame
can be obtained by the inverse unitary transformation $\mathcal{R}^{\dagger
}(\vec{r})$. The result is
\begin{eqnarray}
|\psi (\vec{r})\rangle &\propto &\!\!\!\!\left( \frac{1}{r}-\frac{1}{a_{%
\mathrm{\scriptscriptstyle R}}}\right) |\mathrm{S}\rangle -i\frac{\lambda }{2%
}(\hat{\sigma}_{z1}-\hat{\sigma}_{z2})\cdot \left( \frac{x}{r}\right) |%
\mathrm{S}\rangle \,,  \notag \\
&&\hspace{25mm}\mbox{for }r_{\ast }\lesssim r<<r_{s}.  \label{bps3b}
\end{eqnarray}%
As in the above subsection, Eq.~(\ref{bps3b}) does not depend on whether or
not $|\psi (\vec{r})\rangle $ is an eigenfuction of $H$, and thus is
generally applicable. Finally, we emphasis that the last term in Eq.~(\ref%
{bps3b}) is in the order of unit and cannot be neglected.

Let $|\phi (\vec{r})\rangle $ be the wave function given by the Schr\"{o}%
dinger equation with Hamiltonian $H_{0}$ in Eq.~(\ref{hh}) together with the
\textit{modified} BP boundary condition
\begin{equation}
\lim_{r\rightarrow 0}|\phi (\vec{r})\rangle \propto \left( \frac{1}{r}-\frac{%
1}{a_{\mathrm{\scriptscriptstyle R}}}\right) |\mathrm{S}\rangle -i\frac{%
\lambda }{2}(\hat{\sigma}_{z}^{(1)}-\hat{\sigma}_{z}^{(2)})\cdot \left(
\frac{x}{r}\right) |\mathrm{S}\rangle +\mathcal{O}(r)\,.  \label{bp1d}
\end{equation}%
It is clear that in the whole region $r\gtrsim r_{\ast }$, $|\phi (\vec{r}%
)\rangle $ has the same behavior with the solution of the Schr\"{o}dinger
equation with Hamiltonian $H$ in Eq.~(\ref{hh}). Therefore, theoretical
calculation can be done by replacing $U(\vec{r})$ with condition (\ref{bp1d}).

So far we have obtained the the modified BP boundary condition (\ref{bp1d})
for the system with one-dimensional SO coupling. Comparing Eq.~(\ref{bp1d})
to Eq.~(\ref{bps2}), we find that the SO coupling has two effects on the BP
condition. First, the modified BP condition includes an anisotropic term $%
-i\lambda (\hat{\sigma}_{z}^{(1)}-\hat{\sigma}_{z}^{(2)})x/(2r)|\mathrm{S}%
\rangle $. Second, the scattering length $a_{\mathrm{\scriptscriptstyle R}}$
is determined by the detail of the rotated interaction potential $U_{\mathrm{%
R}}(\vec{r})$ defined in Eq.~(\ref{ur}).

\textit{Relevance to the current experiments.} In general, to obtain the
value of $a_{\mathrm{\scriptscriptstyle R}}$ we need to explicitly solve the
Schr\"{o}dinger equation with potential $U_{\mathrm{R}}(\vec{r})$ for a
given SO coupling. As shown below, however, in the special case of the
current experiments \cite{SOC_Fermi,SOC_MIT}, the SO coupling \textit{does
not} change the scattering length. Namely, $a_{\mathrm{\scriptscriptstyle R}%
} $ equals to the scattering length $a_{0}$ for the case without SO coupling.

In Refs.~\cite{SOC_Fermi,SOC_MIT}, the spin states $|\uparrow \rangle $ and $%
|\downarrow \rangle $ are two hyperfine states of $^{6}$Li or $^{40}$K atom,
and the single-atom Hamiltonian in the Schr\"{o}dinger picture is given by%
\begin{equation}
H_{\mathrm{1bS}}=\frac{\vec{P}^{2}}{2}+\frac{\Omega }{2}\left( \hat{\sigma}%
_{+}e^{2ik_{r}X}+\hat{\sigma}_{-}e^{-2ik_{r}X}\right) +\frac{\delta }{2}\hat{%
\sigma}_{z},
\end{equation}%
with $X$ the single-atom coordinate in the $x$ direction, $\hat{\sigma}%
_{+}=|\uparrow \rangle \langle \downarrow |$ and $\hat{\sigma}_{-}=\hat{%
\sigma}_{+}^{\dagger }$. Here $\Omega $ and $\delta $ are the Rabi frequency
and the two-photon detuning, respectively. Then, in the Schr\"{o}dinger
picture the two-atom Hamiltonian is $H_{\mathrm{2bS}}=H_{\mathrm{1bS}%
}^{(1)}+H_{\mathrm{1bS}}^{(2)}+U_{0}(\vec{r})$, where the bare inter-atomic
interaction potential $U_{0}(\vec{r})$ has the scattering length $a_{0}$.

The SO coupling term emerges after the spin rotation along the $z$-axis or
the unitary transformation $\mathcal{T}(X)=\exp \left[ -ik_{r}X\hat{\sigma}%
_{z}\right] $ is applied. In the transformed picture, the single-atom
Hamiltonian is given by $H_{\mathrm{1b}}=\mathcal{T}(X)H_{\mathrm{1bS}}%
\mathcal{T}^{\dag }(X)$. It can be shown that $H_{\mathrm{1b}}$ takes the
form in Eq. (\ref{h1b2}) with $\lambda =2k_{r}$ and $Z=\Omega \hat{\sigma}%
_{x}/2+\delta \hat{\sigma}_{z}/2$.

The two-atom Hamiltonian in the transformed picture can be written as $H_{%
\mathrm{1b}}^{(1)}+H_{\mathrm{1b}}^{(2)}+U(\vec{r})$, and the transformed
interaction potential $U(\vec{r})$ is given by%
\begin{eqnarray}
U(\vec{r}) &=&\mathcal{T}(X_{1})\mathcal{T}(X_{2})U_{0}(r)\mathcal{T}^{\dag
}(X_{2})\mathcal{T}^{\dag }(X_{2})  \notag \\
&=&e^{-ik_{r}[\hat{\sigma}_{z}^{(1)}-\hat{\sigma}_{z}^{(2)}]x}U_{0}(\vec{r}%
)e^{ik_{r}[\hat{\sigma}_{z}^{(1)}-\hat{\sigma}_{z}^{(2)}]x},  \label{u}
\end{eqnarray}%
where the relative coordinate $x$ satisfies $x=X_{1}-X_{2}$ and we have used the fact $%
[U_{0}(\vec{r}),\hat{\sigma}_{z}^{(1)}+\hat{\sigma}_{z}^{(2)}]=0$, arising
from the conservation of the total $z$-component of hyperfine spin during
the collision process. We emphasis that, in the transformed picture where
the SO-coupling term $\hat{\sigma}_zP_x$ appears, the inter-atomic
interaction is not the bare potentianl $U_{0}(\vec{r})$, but the transformed
one $U(\vec{r})$.

The scattering length $a_{\mathrm{\scriptscriptstyle R}}$ in the modified BP
boundary condition (\ref{bp1d}) is determined by $U_{R}(\vec{r})$ in Eq. (%
\ref{ur}). Substituting Eq. (\ref{u}) into Eq. (\ref{ur}), we find that the
rotated potential $U_{R}(\vec{r})$ reduces to the bare potential--i.e., $%
U_{R}(\vec{r})=U_{0}(r)$. Therefore, the scattering length remains unchanged
with SO coupling--i.e., $a_{\mathrm{\scriptscriptstyle R}}=a_{0}$.

\subsection{With arbitrary type of SO coupling}

We shall now extend the above treatment to the spin-$1/2$ fermonic system
with arbitrary type of SO coupling. In this case, the single-atom
Hamiltonian can be generally written as
\begin{equation}
H_{\mathrm{1b}}=\frac{\vec{P}^{2}}{2}+\lambda \vec{M}\cdot \vec{P}+Z,
\label{h12}
\end{equation}%
with $\vec{M}$ and $Z$ are operators in the spin space and the maximum
eigenvalue of $\vec{M}$ is of the order of unit.
% The term $\lambda \vec{M%
%}\cdot \vec{P}$ describes the SO coupling and $Z$ accounts for the residual
%spin-dependent part. Here,
%and $\lambda $ indicates the intensity of the SO coupling.
%For system discussed in our above section, one has $\vec{M}=(\hat{\sigma}%
%_{z},0,0)$ and $Z=\Omega \hat{\sigma}_{x}/2$.
%and the associated Hamiltonian reads
%for the atomic relative motion is given by%
After being separated from the mass-center motion, the the relative motion
of the two atoms is described by the Hamiltonian
\begin{equation}
H=\vec{p}^{\,2}+\lambda \vec{c}{\cdot \vec{p}}+B(\vec{K})\equiv H_{0}+U({%
\vec{r}}).  \label{h}
\end{equation}%
with $\vec{c}=\vec{M}^{(1)}-\vec{M}^{(2)}$ and $B({\vec{K}}%
)=Z^{(1)}+Z^{(2)}+\lambda ({\vec{M}}^{(1)}+{\vec{M}}^{(2)})\cdot {\vec{K}/2}$%
.

As in the above subsection, to investigate the short-range behavior of the
eigenfunction $|\psi (\vec{r})\rangle $ of $H$, we introduce a unitary
transformation $\mathcal{R}(\vec{r})$ as
\begin{equation}
\mathcal{R}({\vec{r}})=e^{i\lambda c_{x}x/2}e^{i\lambda c_{y}y/2}e^{i\lambda
c_{z}z/2},  \label{ug}
\end{equation}%
with $\vec{c}\equiv (c_{x},c_{y},c_{z})$. The rotated Hamiltonian $H_{%
\mathrm{R}}=\mathcal{R}(\vec{r})H\mathcal{R}^{\dagger }(\vec{r})$ can be
calculated as
\begin{equation}
H_{\mathrm{R}}=\vec{p}^{\,2}-2\lambda \vec{d}(\lambda \vec{r})\cdot \vec{p}%
+W(\vec{r})+U_{\mathrm{R}}(\vec{r})  \label{hrr}
\end{equation}%
with operators $\vec{d}\equiv (d_{x},d_{y},d_{z})$ and $W$ given by
\begin{eqnarray}
d_{x}\left( \lambda \vec{r}\right)  &=&0;  \label{d1} \\
d_{y}\left( \lambda \vec{r}\right)  &=&e^{i\lambda c_{z}z/2}\frac{c_{y}}{2}%
e^{-i\lambda c_{z}z/2}-\mathcal{R}(\vec{r})\frac{c_{y}}{2}\mathcal{R}%
^{\dagger }(\vec{r}); \\
d_{z}\left( \lambda \vec{r}\right)  &=&\frac{c_{z}}{2}-\mathcal{R}(\vec{r})%
\frac{c_{z}}{2}\mathcal{R}^{\dagger }(\vec{r}),  \label{d3}
\end{eqnarray}%
and%
\begin{eqnarray}
W(\vec{r}) &=&i\lambda \left[ \nabla \cdot \vec{d}(\lambda \vec{r})\right] +%
\mathcal{R}(\vec{r})B(\vec{K})\mathcal{R}^{\dagger }(\vec{r})+  \label{w} \\
&&\lambda ^{2}\left[ |\vec{d}(\lambda \vec{r})|^{2}-\mathcal{R}(\vec{r})%
\frac{|\vec{c}|^{2}}{4}\mathcal{R}^{\dagger }(\vec{r})\right] .
\end{eqnarray}%
Here we have $U_{\mathrm{R}}(\vec{r})=\mathcal{R}(\vec{r})U(\vec{r})\mathcal{%
R}^{\dagger }(\vec{r})$ as before.

Unlike Eq.~(\ref{hrs}) in the above subsection, the SO coupling still exists
in the rotated Hamiltonian $H_{\mathrm{R}}$ in Eq.~(\ref{hrr}).
Nevertheless, according to Eqs.~(\ref{d1}-\ref{d3}), we have $\vec{d}%
(\lambda \vec{r})=\mathcal{O}(\lambda r)$. Namely, its zeroth-order
contribution in $H_{\mathrm{R}}$ vanishes. This leads to the following
important property of the eigenfunction $|\psi (\vec{r})\rangle _{\mathrm{R}%
} $ of $H_{\mathrm{R}}$:
\begin{equation}
|\psi (\vec{r})\rangle _{\mathrm{R}}\propto \left( \frac{1}{r}-\frac{1}{a_{%
\mathrm{\scriptscriptstyle R}}}\right) |\mathrm{S}\rangle \hspace{5mm}(%
\mbox{for }r_{\ast }\lesssim r<<r_{s})\,.  \label{kaka}
\end{equation}

The result in Eq.~(\ref{kaka}) is proved as follows. In the region $%
r>r_{\ast }$ where $U_{R}(\vec{r})$ is negligible, the eigen-equation of $H_{%
\mathrm{R}}$ reads%
\begin{equation}
\left[ {\vec{p}}^{2}-2\lambda \vec{d}(\lambda \vec{r})\cdot \vec{p}+W(\vec{r}%
)\right] |\psi (\vec{r})\rangle _{\mathrm{R}}=E|\psi (\vec{r})\rangle _{%
\mathrm{R}}.  \label{a1}
\end{equation}%
As shown in appendix B, in this region the wave function $|\psi (\vec{r}%
)\rangle _{\mathrm{R}}$ can be expressed as
\begin{eqnarray}
&&|\psi (\vec{r})\rangle _{\mathrm{R}}=\frac{C_{-1}}{r}|\mathrm{S}\rangle
+\sum_{n=0}^{\infty }C_{n}r^{n}|\mathrm{S}\rangle  \notag \\
&&+\sum_{l=1}^{\infty }\sum_{m_{l}=-l}^{l}\sum_{n=0}^{\infty
}r^{n}Y_{l,m_{l}}(\theta ,\phi )|A_{l,m_{l},n}\rangle ,  \label{a2}
\end{eqnarray}
in the spherical coordinate $\left( r,\theta ,\phi \right) $. Here $%
Y_{l,m_{l}}(\theta ,\phi )$ are the spherical harmonic functions, $C_{n}$ $%
(n=-1,0,1,...)$ is the coefficient of term $r^{n}|\mathrm{S}\rangle $, and $%
|A_{l,m_{l},n}\rangle $ $(n=0,1,...)$ is the spin-state with respect to $%
r^{n}Y_{l,m_{l}}(\theta ,\phi )$. Substituting Eq. (\ref{a2}) into Eq. (\ref%
{a1}) and comparing the coefficient of the term $r^{-2}$ in both sides, we
find that because $d(\lambda \vec{r})=\mathcal{O}(\lambda r)$, one has $%
|A_{l,m_{l},0}\rangle =0.$ Therefore, in the short-range region $|\psi(\vec{r%
})\rangle_{\mathrm{R}}$ behaves as in Eq. (\ref{kaka}), with the scattering
length $a_{\mathrm{\scriptscriptstyle R}}$ determined by both the potential $%
U_{\mathrm{R}}(\vec{r})$ and the operator $\vec{d}(\lambda \vec{r})$.

With Eq.~(\ref{kaka}) and following the procedure in the above subsection,
we obtain %the wave function $|\psi (\vec{r})\rangle $ as
%\begin{eqnarray}
%|\psi (\vec{r})\rangle &=&\mathcal{U}^{\dagger }(\vec{r})|\psi (\vec{r}%
%)\rangle _{\mathrm{R}}  \notag \\
%&\propto &\left( \frac{1}{r}-\frac{1}{a}\right) |\mathrm{S}\rangle -i\frac{%
%\lambda }{2}\vec{c}\cdot \left( \frac{\vec{r}}{r}\right) |\mathrm{S}\rangle \, ,
%\label{bps3}
%\end{eqnarray}%
%as well as
the \textit{modified} BP boundary condition for the general type of SO
coupling as
\begin{equation}
\lim_{r\rightarrow 0}|\phi (\vec{r})\rangle \propto \left( \frac{1}{r}-\frac{%
1}{a_{\mathrm{\scriptscriptstyle R}}}\right) |\mathrm{S}\rangle -i\frac{%
\lambda }{2}\vec{c}\cdot \left( \frac{\vec{r}}{r}\right) |\mathrm{S}\rangle +%
\mathcal{O}(r).  \label{rbps}
\end{equation}
As in the above subsection, the value of the scattering length $a_{\mathrm{%
\scriptscriptstyle R}}$ in general depends on the SO coupling. This
dependence is also shown in Fig.~3 of Ref.~\cite{fewbody1} with a simple 
 model where the potential $U(\vec{r})$ is modeled as a 
 spin-independent spherical square well.

\section{Modified BP boundary condition for atoms with arbitrary spin}

Finally, we consider the general case: a system of two fermonic or bosonic
atoms with any kind of SO coupling and arbitrary spin. The Hamiltonian for
the single-atom motion and the relative motion of the two atoms are still
given by Eqs. (\ref{h12}) and (\ref{h}), respectively. %Now $\vec{M}$
%and $Z$ are operators in the Hilbert space of the spin of the two atoms. In
%this section

For simplicity, we first consider the case that the inter-atomic interaction
$U(\vec{r})$ (with scattering length $a$) is independent of the atomic spin.
In this case, it can be shown that without SO coupling, the low-energy
eigen-state $|\psi (\vec{r})\rangle $ of the relative-motion Hamiltonian $%
(-\nabla ^{2}+U(\vec{r})+Z^{(1)}+Z^{(2)})$ behaves as
\begin{equation}
|\psi (\vec{r})\rangle \propto \left( \frac{1}{r}-\frac{1}{a}\right) |\chi
\rangle ,\hspace{5mm}\mbox{for }r_{\ast }\lesssim r<<r_{s}\;.
\label{general:psi}
\end{equation}%
%
%
%
%
%
%
%
%
%
%
%
%
%
%
%when the relevant eigen-energy is much smaller than $1/r_{\ast }^{2}$.
%Equation~(\ref{general:psi})
This is very similar to Eq.~(\ref{2}), but now the $\vec{r}$-independent
spin state $|\chi \rangle $ is not unique. Instead, $|\chi \rangle $ can be
different for different eigen-states $|\psi (\vec{r})\rangle $.

%Here $%
%a$ is the scattering length with respect to $U(r)$,
%%and $|\chi \rangle $ is
%a $\vec{r}$-independent state of atomic spin. The spin state $|\chi \rangle $
%with respect to  can be
%different.
In the presence of SO coupling, the short-range behavior of the
eigenfunction $|\psi (\vec{r})\rangle $ can be obtained via the same
approach using the unitary transformation $\mathcal{R}({\vec{r}})$ in Eq.~(%
\ref{ug}). In particular, in the region $r<<r_{s}$, it is sufficient to keep
the lowest-order terms of $\vec{d}(\lambda \vec{r})$ and $W(\vec{r})$
defined in Eqs. (\ref{d1}-\ref{w}). Thus, the rotated wave function $|\psi (%
\vec{r})\rangle _{\mathrm{R}}=\mathcal{R}(\vec{r})|\psi (\vec{r})\rangle $
satisfies the equation%
\begin{eqnarray}
\left[ \vec{p}^{\,2}-2\lambda \vec{g}(\lambda \vec{r})\cdot \vec{p}+W(0)+U(%
\vec{r})\right] |\psi (\vec{r})\rangle _{\mathrm{R}} &=&E|\psi (\vec{r}%
)\rangle _{\mathrm{R}},  \notag \\
(r &<&<r_{s})  \label{a31}
\end{eqnarray}%
where $E$ is the eigen-energy. Here we have used $\vec{d}(0)=0$ and the fact 
$U_{\mathrm{R}}(\vec{r})=U(\vec{r})$ which is because $U$ is
spin-independent. The operator $\vec{g}=(g_{x},g_{y},g_{z})$ is defined as $%
g_{i}(\lambda \vec{r})=\vec{r}\cdot \lbrack \nabla d_{i}(\lambda \vec{r})|_{%
\vec{r}=0}]$ with $i=x,y,z$. In Eq. (\ref{a31}) the term $-2\lambda \vec{g}%
\cdot \vec{p}$ couples the $s$-wave and $d$-wave components of $|\psi (\vec{r%
})\rangle _{\mathrm{R}}$. In the Schoredinger equation, the coupling terms
are either independent of $r$ or proportional to $r(\partial /\partial r)$, and thus
do not decrease the power of $r$ in the wave function $|\psi (\vec{r}%
)\rangle _{\mathrm{R}}$. The estimation with the semi-classical approximation 
$\partial |\psi (\vec{r})\rangle _{%
\mathrm{R}}/\partial r\lesssim \sqrt{-U(\vec{r})}|\psi (\vec{r})\rangle _{%
\mathrm{R}}$ shows that, in the
short-range region $r_{\ast }\lesssim r<<r_{s}$, the intensity of this
coupling is much smaller than the centrifugal potential $6/r^{2}$, which is
the energy gap between the $s$-wave and $d$-wave channels. 
In addition, for many systems this intensity is
also much smaller than $6/r^{2}$ even when $r\lesssim r_{\ast }$. An example
is a system with Lennard-Jones potential $U(\vec{r}%
)=-c_{6}/r^{6}+c_{12}/r^{12}$. Therefore, for these systems we can neglect
the SO coupling in the entire region $r<<r_{s}$. Then one has
\begin{equation}
|\psi (\vec{r})\rangle _{\mathrm{R}}\propto \left( \frac{1}{r}-\frac{1}{a}%
\right) |\chi \rangle \hspace{5mm}(\mbox{for }r_{\ast }\lesssim r<<r_{s}).
\end{equation}%
Note that $a$ is still the scattering length of the potential $U(r)$.
Accordingly, we have the modified BP boundary condition
\begin{equation}
\lim_{r\rightarrow 0}|\phi (\vec{r})\rangle \propto \left( \frac{1}{r}-\frac{%
1}{a}\right) |\chi \rangle -i\frac{\lambda }{2}\vec{c}\cdot \left( \frac{%
\vec{r}}{r}\right) |\chi \rangle +\mathcal{O}(r).  \label{lla}
\end{equation}

The situation becomes more sophisticated if $U(\vec{r})$ is spin-dependent
or the SO coupling cannot be neglected in $H_R$ when $r\lesssim r_{\ast }$. In these
cases, $1/a$ in the modified BP boundary condition (\ref{lla}) should be
replaced by an operator $A_{\mathrm{R}}$ in the spin space, which is also
determined by $U_{\mathrm{R}}(\vec{r})$ and $\vec{d}(\lambda \vec{r})$. The
detail is given in appendix C.

We conclude this section by pointing out that as in Sec.~II, in the current
experiments~\cite%
{NIST,NIST_elec,SOC,collective_SOC,JingPRA,NIST_partial,ourdecay} for
bosonic atoms with one-dimensional SO coupling, the rotated potential $U_{%
\mathrm{R}}$ is equivalent to the bare potential $U_{0}$ in the Schr\"{o}%
dinger picture, and the operator $\vec{d}$ is zero. Then the operator $A_{%
\mathrm{R}}$ in the modified BP boundary condition is independent of the SO
coupling. For instance, for the ultracold gases with spin-$1$ $^{87}$Rb
atoms, we have $A_{\mathrm{R}}=1/a_{0}\mathcal{P}_{F=0}+1/a_{2}\mathcal{P}%
_{F=2}$, where $a_{0}$ $(a_{2})$ is the scattering length with respect to
the total atomic spin $F=0$ ($F=2$) and $\mathcal{P}_{F=0,2}$ are the
relevant projection operators.

\section{Discussion}

In this paper we derive the modified BP boundary condition for ultracold
atomic gases with SO coupling. It is shown that the SO coupling brings a new
anisotropic term to the BP boundary condition, and may change the value of
atomic scattering length.

Our result can be used for the research of both few-body and many-body
problems in SO-coupled ultracold gases. For instance, for $N$ spin-$1/2$
fermonic atoms with the Hamiltonian
\begin{eqnarray}
H_{T} &=&\sum_{i=1}^{N}H_{\mathrm{1b}}(i)+\sum_{i=1}^{N}V_{\mathrm{trap}%
}^{(i)}+\sum_{i<j}^{N}U({\vec{r}}_{ij})  \notag \\
&\equiv &H_{F}+\sum_{i<j}^{N}U({\vec{r}}_{ij}),  \label{mb2}
\end{eqnarray}%
one can replace the interaction potential $U({\vec{r}}_{ij})$ by the
modified BP boundary condition
\begin{eqnarray}
&&\lim_{|{\vec{r}}_{ij}\mathbf{|}\rightarrow 0}\langle \vec{r}_{ij}|\Phi
\rangle  \notag \\
&\propto &\left[ \left( \frac{1}{|{\vec{r}}_{ij}\mathbf{|}}-\frac{1}{a_{%
\mathrm{\scriptscriptstyle R}}}\right) |\mathrm{S}\rangle _{ij}-i\frac{%
\lambda }{2}\vec{c}\cdot \left( \frac{\vec{r}_{ij}}{|{\vec{r}}_{ij}\mathbf{|}%
}\right) |\mathrm{S}\rangle _{ij}\right] |\Phi ^{\prime }\rangle  \notag \\
&&+\mathcal{O}(r_{ij}).  \label{bpc1}
\end{eqnarray}%
In Eq. (\ref{mb2}) $V_{\mathrm{trap}}^{(i)}$ is the trap potential for the $%
i $th atom and ${\vec{r}}_{ij}$ is the relative position between the $i$th
and $j$th atoms; in Eq. (\ref{bpc1}) $|\Phi \rangle $ is the $N$-atom state,
$|\vec{r}_{ij}\rangle $ is the eigen-state of the relative motion of the ($%
i,j$) pair, $|\mathrm{S}\rangle _{ij}$ is the singlet spin state for the two
atoms and $|\Phi ^{\prime }\rangle $ is a quantum state for other atoms. The
limit in Eq. (\ref{bpc1}) is taken for fixing the positions of other atoms
as well as the mass center of the $\left( i,j\right) $ pair. In the region $|%
{\vec{r}}_{ij}\mathbf{|}\gtrsim r_{\ast }$, the solution of the Schr\"{o}%
dinger equation with the free Hamiltonian $H_{F}$ under the boundary
condition (\ref{bpc1}) has the same behavior as the soulution of the Schr%
\"{o}dinger equation with the total Hamiltonian $H_{T}$.

Our result is also useful in the current experiments of ultracold gases with
one-dimensional SO coupling and far away from the Feshbach resonance point.
As shown in Secs. II and III, the scattering lengths in these systems are
independent of the SO coupling, and thus all the terms in the modified BP
boundary condition can be fully determined with the known parameters.

\begin{acknowledgments}
We thank H. Zhai, X. Chui, H. Duan and L. You for useful discussions. This work is supported by National Natural Science Foundation of China under
Grants No. 11074305, 11222430, 11275185, 10975127, CAS, NKBRSF of China under Grants No.
2012CB922104, 2011CB921300 and the Research Funds of Renmin University of China
(10XNL016). 

\end{acknowledgments}

\appendix%\appendixpage
\addcontentsline{toc}{section}{Appendices}\markboth{APPENDICES}{}
\begin{subappendices}

\section{Short-range behavior of scattering wave function}

In this appendix we prove Eq. (\ref{bps}) for the short-range behavior of
the wave function $|\psi (\vec{r})\rangle $ in the cases without SO
coupling. Without loss of generality, here we consider the case that $|\psi (%
\vec{r})\rangle $ is the scattering wave function and then satisfies the
Lippmman-Schwinger equation%
\begin{equation}
|\psi (\vec{r})\rangle =|\psi ^{(0)}(\vec{r})\rangle +\int d\vec{r}%
^{\,\prime }g_{0}(E,\vec{r},\vec{r}^{\,\prime })U(\vec{r}^{\,\prime })|\psi (%
\vec{r}^{\,\prime })\rangle ,  \label{bb1}
\end{equation}%
Here $E$ is the eigen-energy of $H$ in Eq. (\ref{2}) with respect to $|\psi (%
\vec{r})\rangle $, $|\psi ^{(0)}(\vec{r})\rangle $ is the incident state and
satisfies $(\vec{p}^{\,2}+B)|\psi ^{(0)}(\vec{r})\rangle =E|\psi ^{(0)}(\vec{%
r})\rangle $. For our system with two fermonic atoms, $|\psi ^{(0)}(\vec{r}%
)\rangle $ is anti-symnetric with respect to the permutation of the two
atoms. In Eq. (\ref{bb1}) the Green's operator $g_{0}(E,\vec{r},\vec{r}%
^{\prime })$ is defined as%
\begin{eqnarray}
g_{0}(\eta ,\vec{r},\vec{r}^{\,\prime }) &=&\frac{1}{\eta +i0^{+}-\left(
\vec{p}^{\,2}+B\right) }\delta (\vec{r}-\vec{r}^{\,\prime })  \notag \\
&=&-\sum_{n}\frac{e^{i\sqrt{\eta -\varepsilon _{n}}\left\vert \vec{r}-\vec{r}%
^{\prime }\right\vert }}{\pi \left\vert \vec{r}-\vec{r}^{\,\prime
}\right\vert }|n\rangle \langle n|  \label{gr}
\end{eqnarray}%
with $\varepsilon _{n}$ and $|n\rangle $ the $n$th eigen-value and
eigen-state of the operator $B$, respectively.

Since the potential $U(\vec{r})$ is negligible in the region $r>r_{\ast }$,
the integration in Eq.~(\ref{bb1}) is only effective in the region $%
r^{\prime }\leq r_{\ast }$. In the low-energy cases, when $r\rightarrow
\infty $ and $r^{\prime }\leq r_{\ast }$, the function $g_{0}(E,\vec{r},\vec{%
r}^{\,\prime })$ becomes very steady with respect to $\vec{r}^{\,\prime }$
and we have $g_{0}(E,\vec{r},\vec{r}^{\,\prime })\approx g_{0}(E,\vec{r},0)$%
. Therefore, in the limit $r\rightarrow \infty $, the solution of Eq.~(\ref%
{bb1}) takes the form
\begin{equation}
|\psi (\vec{r})\rangle =|\psi ^{(0)}(\vec{r})\rangle +g_{0}(E,\vec{r}%
,0)|\chi \rangle ,  \label{bb4}
\end{equation}%
where the spin state $|\chi \rangle $ is related to $|\psi (\vec{r})\rangle $
via the equation $|\chi \rangle =\int d\vec{r}^{\,\prime }U(\vec{r}%
^{\,\prime })|\psi (\vec{r}^{\,\prime })\rangle $. Furthermore, due to the
facts $P_{12}|\psi (\vec{r})\rangle =-|\psi (\vec{r})\rangle $ and $P_{12}U(%
\vec{r})P_{12}=U(\vec{r})$ with $P_{12}$ the permutation operator of the two
atoms, one finds that $P_{12}U(\vec{r})|\psi (\vec{r})\rangle =-U(\vec{r}%
)|\psi (\vec{r})\rangle $. This result yields
\begin{equation}
|\chi \rangle =|\mathrm{S}\rangle \int d\vec{r}^{\,\prime }U\!\left(
r^{\prime }\right) \langle \mathrm{S}|\psi (\vec{r}^{\,\prime })\rangle .
\label{bb5}
\end{equation}

On the other hand, since $|\psi (\vec{r})\rangle $ is an eigen-state of $H$
and the potential $U(\vec{r})$ is negligible in the region $r>r_{\ast }$, in
such a region the wave function $|\psi (\vec{r})\rangle $ satisfies the
equation
\begin{equation}
\left( \vec{p}^{\,2}+B\right) |\psi (\vec{r})\rangle =E|\psi (\vec{r}%
)\rangle .  \label{bb3}
\end{equation}%
Therefore, the behavior of the wave function $|\psi (\vec{r})\rangle $ in
the region $r\gtrsim r_{\ast }$ is determined by Eq.~(\ref{bb3}) and the
boundary condition (\ref{bb4}) in the limit $r\rightarrow \infty $.
Considering Eq. (\ref{bb5}), one can easily prove that the function
\begin{equation}
|\psi (\vec{r})\rangle =|\psi ^{(0)}(\vec{r})\rangle +\Lambda _{0}g_{0}(E,%
\vec{r},0)|\mathrm{S}\rangle  \label{bb6}
\end{equation}%
with $\Lambda _{0}$ a constant satisfies both of the two conditions. Therefore, $|\psi (\vec{r})\rangle $
satisfies Eq.~(\ref{bb6}) in the whole region of $r\gtrsim r_{\ast }$.

To obtain the short-range behavior of $|\psi (\vec{r})\rangle $, one can
expand Eq. (\ref{bb6}) as a series of $r$, and then neglect the high-order
terms. Using Eq. (\ref{gr}) and the fact that $P_{12}|\psi ^{(0)}(\vec{r}%
)\rangle =-|\psi ^{(0)}(\vec{r})\rangle $, we immediately gets the result in
Eq. (\ref{bps}):
\begin{equation}
|\psi (\vec{r})\rangle \propto \left( \frac{1}{r}-\frac{1}{a}\right) |%
\mathrm{S}\rangle \hspace{5mm}\mbox{for }r_{\ast }\lesssim r<<1/\sqrt{%
\varepsilon }\,.
\end{equation}

\section{Proof of Eq. (\ref{a2})}

Now we prove Eq. (\ref{a2}) for the behavior of $|\psi (\vec{r})\rangle _{%
\mathrm{R}}$ in the region $r\gtrsim r_{\ast }$. To this end, we first
consider the behavior of the un-rotated eigenfunction $|\psi (\vec{r}%
)\rangle $ of Hamiltonian $H$ defined in Eq. (\ref{h}). Without loss of
generality, here we consider the case that $|\psi (\vec{r})\rangle $ is the
scattering wave function. Using the approach in appendix A, we can prove
that when $r\gtrsim r_{\ast }$ we have%
\begin{equation}
|\psi (\vec{r})\rangle =|\psi ^{(0)}(\vec{r})\rangle +\Lambda_0g(E ,\vec{r}%
,0)|\mathrm{S}\rangle\,.  \label{b6}
\end{equation}%
Here $E$ is the eigen-energy of $H$ with respect to $|\psi (\vec{r%
})\rangle $, $|\psi ^{(0)}(\vec{r})\rangle $ is the incident state and
satisfies $H_{0}|\psi ^{(0)}(\vec{r})\rangle =E|\psi ^{(0)}(\vec{r%
})\rangle $ with $H_{0}$ defined in Eq. (\ref{h}). The Green's operator $%
g(E ,\vec{r},\vec{r}^{\,\prime })$ is defined as%
\begin{equation}
g(\eta ,\vec{r},\vec{r}^{\,\prime })=\frac{1}{\eta +i0^{+}-H_{0}}\delta (%
\vec{r}-\vec{r}^{\,\prime }).
\end{equation}

Now we expand the r.h.s. of Eq. (\ref{b6}) as a power series of $r$. To this
end, we first consider the operator $F(\vec{k})\equiv \lambda \vec{c}\cdot
\vec{k}+B(\vec{K})$, with $\vec{k}$ a constant operator and $\lambda ,\vec{c}
$ and $B(\vec{K})$ defined in Sec. II. For each given vector $\vec{k}$, $F(%
\vec{k})$ is an operator in the $4$-dimensional spin space. We denote the $%
\alpha $-th ($\alpha =1,2,3,4$) eigen-energy and eigen-state $F(\vec{k})$ as
$\mathcal{E}(\alpha ,\vec{k})$ and $|\alpha (\vec{k})\rangle $,
respectively. Therefore, the incident wave function $|\psi ^{(0)}(\vec{r}%
)\rangle $, which is an eigenfunction of $H_{0}$ defined in Eq. (\ref{h}),
takes the form%
\begin{equation}
|\psi ^{(0)}(\vec{r})\rangle =\frac{1}{2(2\pi )^{3/2}}(1-P_{12})e^{i\vec{k}%
\cdot \vec{r}}|\alpha (\vec{k})\rangle .  \label{b7}
\end{equation}%
with $P_{12}$ the permutation operator for both the spin and the spatial
motion of the two atoms. Eq. (\ref{b7}) leads to the result that
\begin{equation}
|\psi ^{(0)}(\vec{r})\rangle =\mathcal{O}(r^{0}).  \label{b11}
\end{equation}

Now we consider the expansion of the Green's function $g(E ,\vec{r}%
,0)$. Using the fact%
\begin{equation}
\delta \left( \vec{r}-\vec{r}^{\,\prime }\right) =\int d\vec{k}\frac{e^{i%
\vec{k}\cdot \left( \vec{r}-\vec{r}^{\,\prime }\right) }}{\left( 2\pi
\right) ^{3}}\left( \sum_{\alpha }|\alpha (\vec{k})\rangle \langle \alpha (%
\vec{k})|\right) ,
\end{equation}%
it is easy to show that
\begin{equation}
g(E ,\vec{r},0)=\sum_{\alpha }\int d\vec{k}\frac{e^{i\vec{k}\cdot
\vec{r}}}{\left( 2\pi \right) ^{3}}\frac{|\alpha (\vec{k})\rangle \langle
\alpha (\vec{k})|}{E +i0^{+}-\left[ \vec{k}^{2}+\mathcal{E}(\alpha
,\vec{k})\right] }.  \label{b10}
\end{equation}%
Eq. (\ref{b10}) and the completeness relationship $\sum_{\alpha }|\alpha (%
\vec{k})\rangle \langle \alpha (\vec{k})|=1$ lead to the result%
\begin{eqnarray}
&&g(E ,\vec{r},0)=\int d\vec{k}\frac{e^{i\vec{k}\cdot \vec{r}}}{%
\left( 2\pi \right) ^{3}}\frac{1}{E +i0^{+}-\vec{k}^{2}}  \notag
\label{b12} \\
&&+\sum_{\alpha }\int d\vec{k}\frac{e^{i\vec{k}\cdot \vec{r}}}{\left( 2\pi
\right) ^{3}}|\alpha (\vec{k})\rangle \langle \alpha (\vec{k})|\times
\notag \\
&&\left( \frac{1}{E +i0^{+}-\left[ \vec{k}^{2}+\mathcal{E}(\alpha ,%
\vec{k})\right] }-\frac{1}{E +i0^{+}-\vec{k}^{2}}\right) .  \notag
\\
&&
\end{eqnarray}%
It is pointed out that, in the limit $r\rightarrow 0$, the integration in
the r.h.s of Eq. (\ref{b12}) converges to a constant operator in the spin
space. On the other hand, we also have%
\begin{equation}
\int d\vec{k}\frac{e^{i\vec{k}\cdot \vec{r}}}{\left( 2\pi \right) ^{3}}\frac{%
1}{E +i0^{+}-\vec{k}^{2}}=-\frac{e^{i\sqrt{E }r}}{\pi
r}.
\end{equation}%
Due to these facts, we have $g(E ,\vec{r},0)\propto 1/r+\mathcal{O}%
(r^{0}).$ Substituting this result and Eq. (\ref{b11}) into Eq. (\ref{b6})
and using the relation $|\psi (\vec{r})\rangle _{\mathrm{R}}=\mathcal{R}(%
\vec{r})|\psi (\vec{r})\rangle $ with $\mathcal{R}(\vec{r})$ defined in Eq. (%
\ref{ug}), we can find that
\begin{equation}
|\psi (\vec{r})\rangle _{\mathrm{R}}\propto \frac{1}{r}|\mathrm{S}\rangle +%
\mathcal{O}(r^{0}),
\end{equation}%
Namely, $|\psi (\vec{r})\rangle _{\mathrm{R}}$ takes the form of Eq. (\ref%
{a2}).

\section{The modified BP boundary condition for atoms with spin-dependent
interaction}

In Sec. III of our maintext, we derive the modified BP boundary condition
for atoms with arbitrary spin and SO coupling. Our result in Eq.~(\ref{lla})
is based on
the following two assumptions: (a) the inter-atomic interaction is spin-independent;
(b) in the rotated frame, the influence of the SO coupling or the term
$-2\vec{g}\cdot\vec{p}$ is negligible in the region $r\lesssim r_{\ast}$.
In this appendix we go beyond these two assumptions and
derive the general type of the modified BP boundary condition
for atoms with with arbitrary spin and SO coupling.

We first go beyond the assumption (a) and consider the case
of atoms with spin-dependent interaction $U(\vec{r})$. When there is no SO
coupling, the eigenfunction $|\psi (\vec{r})\rangle $ of the two-atom
relative Hamiltonian satisfies
\begin{equation}
\left[ -\nabla ^{2}+U(\vec{r})+Z^{(1)}+Z^{(2)}\right] |\psi (\vec{r})\rangle
=E|\psi (\vec{r})\rangle .  \label{c1}
\end{equation}%
We assume the spin space of the two atoms is $n$ dimensional, and denote the
eigen-states of $Z^{(1)}+Z^{(2)}$ as $|j\rangle $ ($j=1,...,n$). We further
define
\begin{equation*}
|\Psi (\vec{r})\rangle =r|\psi (\vec{r})\rangle \,.
\end{equation*}%
Then $|\Psi (\vec{r})\rangle $ satisfies the boundary condition $|\Psi
(0)\rangle =0$.

We first consider that $U(\vec{r})$ is spherical. Thus, $|\Psi (\vec{r})\rangle $
can be written as
\begin{equation}
|\Psi (\vec{r})\rangle =\sum_{j=1}^{n}\Psi _{j}(r)|j\rangle \,.  \label{c2}
\end{equation}%
Then, Eq. (\ref{c1}) can be re-expressed as the equation for $|\Psi (%
\vec{r})\rangle $. We define $|\Phi ^{(\alpha )}(r)\rangle $ as the $s$-wave
solution of this equation, with component $\Phi _{j}^{(\alpha )}(r)$
satisfying the boundary conditions $\Phi _{j}^{(\alpha )}(0)=0$ and
\begin{equation}
\left. \frac{d}{dr}\Phi _{j}^{(\alpha )}(r)\right\vert _{r=0}=\left\{
\begin{array}{c}
1,\ {\rm for}\ \alpha =j \\
0,\ {\rm for}\ \alpha\neq j%
\end{array}%
\right. .
\end{equation}%
Therefore, the states $|\phi ^{(\alpha )}(\vec{r})\rangle =|\Phi ^{(\alpha
)}(\vec{r})\rangle /r$ are $n$ special solutions of Eq. (\ref{c2}). In the
short-range region, the low-energy wave function $|\phi ^{(\alpha )}(\vec{r}%
)\rangle $ behaves as%
\begin{equation}
|\phi ^{(\alpha )}(\vec{r})\rangle =\frac{1}{r}|M_{\alpha }\rangle
-|T_{\alpha }\rangle
\end{equation}%
with $|M_{\alpha }\rangle $ and $|T_{\alpha }\rangle $ are states of atomic spin.

Furthermore, any $s$-wave solution $|\psi (\vec{r})\rangle $ of Eq. (\ref{c1}%
) can be written as the linear combination of $|\phi ^{(\alpha)}(\vec{r}%
)\rangle $, and then expressed as
\begin{equation}
|\psi (\vec{r})\rangle =\sum_{\alpha=1}^{n}b_{\alpha}\left[ \frac{1}{r}%
|M_{\alpha}\rangle -|T_{\alpha}\rangle \right] \hspace{5mm}\mbox{for }%
r_{\ast }\lesssim r<<r_{s}.  \label{bb}
\end{equation}%
In addition, the low-energy solutions of Eq.~(\ref%
{c1}) with high partial waves are negligible in the short-range region.
Therefore, Eq. (\ref{bb}) is actually satisfied by all the low-energy
solutions of Eq. (\ref{c1}).

When the states $|M_{\alpha }\rangle $ with different $\alpha $ are linearly
independent of each other, we can define an operator $A$ which satisfies $%
A|M_{\alpha }\rangle =|T_{\alpha }\rangle $ (in particular, when the
interaction $U$ is independent of the atomic spin, we have $A=1/a$). With
this definition, the behavior (\ref{bb}) of $|\psi (\vec{r})\rangle $ can be
re-written as
\begin{equation}
|\psi (\vec{r})\rangle \propto \left( \frac{1}{r}-A\right) |\chi \rangle %
\hspace{5mm}\mbox{for }r_{\ast }\lesssim r<<r_{s}.  \label{cc}
\end{equation}%
As in Sec.III, the $\vec{r}$-independent state $|\chi \rangle $ in the
spin space is not unique. Finally, it can be proved that in the low-energy
limit the above result is also correct when $U(\vec{r})$ becomes anisotropic.

In the presence of SO coupling, with Eq.~(\ref{cc}) and the approach in our
maintext we can obtain the modified BP boundary condition:
\begin{equation}
\lim_{r\rightarrow 0}|\phi (\vec{r})\rangle \propto \left( \frac{1}{r}-A_{%
\mathrm{R}}\right) |\chi \rangle -i\frac{\lambda }{2}\vec{c}\cdot \left(
\frac{\vec{r}}{r}\right) |\chi \rangle +\mathcal{O}(r)\,  \label{lls}
\end{equation}%
with the operator $A_{\mathrm{R}}$ determined by both the
potential $U_{\mathrm{R}}(\vec{r})$ and the operator $\vec{d}(\lambda \vec{r}%
)$.

Finally, if we go beyond the assumption (b) and
consider the case that the SO coupling cannot be neglected
when $r\lesssim r_{\ast}$,
we can also follow the above approach, and obtain
 the modified BP boundary condition which has the form in Eq.~(\ref{lls}).

\end{subappendices}

\end{document}